%% file: UnitaryPermutations-arxiv.tex
\documentclass[11pt]{article}
\usepackage{tikz-cd}
\usetikzlibrary{decorations.pathreplacing}
\usepackage{amsmath,amssymb}
\usepackage{authblk}

\input{myQcircuit}

\newcommand{\upn}{$\textsf{UP}_n$ }
\newcommand{\QS}{\texttt{QS}}

\newtheorem{theorem}{Theorem}[section]
\newtheorem{lemma}[theorem]{Lemma}

\newtheorem{corollary}[theorem]{Corollary}
\newtheorem{conjecture}[theorem]{Conjecture}
\newtheorem{definition}[theorem]{Definition}

\begin{document}

\def\titlerunning{Quantum Circuits for the Unitary~Permutation~Problem}
\def\authorrunning{Stefano Facchini and Simon Perdrix}
\title{Quantum Circuits for the Unitary~Permutation~Problem}
\author[1]{Stefano Facchini}
\affil[1]{Univ. Grenoble Alpes, LIG, F-38000 Grenoble, France, \texttt{stefano.facchini@imag.fr}} 
\author[2]{Simon Perdrix}
\affil[1]{CNRS, LORIA, Carte Team, Univ. de Lorraine, Nancy, France \texttt{simon.perdrix@loria.fr}}

\date{}
\maketitle

\begin{abstract}

We consider the \emph{Unitary Permutation} problem which consists, given $n$ unitary gates $U_1, \ldots, U_n$ and a permutation $\sigma$ of $\{1,\ldots, n\}$, in applying the unitary gates in the order specified by $\sigma$, i.e. in performing $U_{\sigma(n)}\circ \ldots \circ U_{\sigma(1)}$.

This problem has been introduced and investigated in \cite{Colnaghi} where two models of computations  are considered. This first is the (standard) model of query complexity: the complexity measure is  the number of calls  to any of the unitary gates  $U_i$ in a quantum circuit which solves the problem.  The second model provides \emph{quantum switches}  and treats unitary transformations as inputs of second order. In that case the complexity measure is the number of quantum switches. In their paper, Colnaghi et al.~\cite{Colnaghi} have shown that the problem can be solved within $n^2$ calls in the query model and $\frac{n(n-1)}2$ quantum switches in the new model.

We refine these results by proving that $n\log_2(n) +\Theta(n)$ quantum switches are necessary and sufficient to solve this problem, whereas $n^2-2n+4$ calls are sufficient to solve this problem in the standard quantum circuit model. We prove, with an additional assumption on the family of gates used in the circuits, that $n^2-o(n^{7/4+\epsilon})$ queries are required, for any $\epsilon >0$. The upper and lower bounds for the standard quantum circuit model are established by pointing out connections with the \emph{permutation as substring} problem introduced by Karp. 

\end{abstract}

\section{Introduction}

The problem of applying two unitary gates $U$ and $V$ in an order specified by a control bit $x$ is a natural problem: one wants to apply $VU$ if $x=0$ and $UV$ if $x=1$. Surprisingly, Chiribella et al.~\cite{Chiribella} showed that in the standard model of quantum circuits, this task cannot be realised using a single call to $U$ and a single call to $V$, whereas, in the lab,  a simple procedure can be implemented -- using standard tools in quantum optics for instance -- that performs this task using a single call to $U$ and and single call to $V$ (see \cite{Chiribella} for details). To model such procedure, they introduced the notion of quantum switch (\QS) which is a gate that inputs two unitary transformations and a control bit and performs a switch or not of the unitary transformations depending on the control qubit: $\QS (x,U_0,U_1) = (U_x, U_{\bar x})$. 

To point out a separation between the standard model of quantum circuits and the quantum switch model, Colnaghi et al.~\cite{Colnaghi} considered the generalisation of the previous problem: given $n$ unitary transformations $U_1,\ldots, $ $U_n$, and a permutation $\sigma$ of $[n]$, the task consists in performing $U_{\sigma_(n)}\circ \ldots U_{\sigma(1)}$, where $[n]$ denotes the interval $\{1,\ldots,n\}$. This problem is called the \emph{Unitary Permutation} problem of size $n$ or $\textsf{UP}_n$ problem for short.  
They proved that the \upn problem can be solved within $n^2$ queries in the standard model of quantum circuits whereas $\frac{n(n-1)}2$ quantum switches suffice to solves this problem. Moreover, they claimed the optimality of their constructions in both cases. However,  we improve both constructions: in the standard model of quantum circuits we show that $n^2-2n+4$ queries are sufficient. To this end we reduce the problem to the existence of a \emph{complete} sequence over the set $\{1,\ldots, n\}$, i.e. a sequence which contains all permutations of $[n]$ as subsequences.  This problem has been originally introduced by Karp \cite[Problem 36]{chvataletal} and bounds on the size of the minimal complete sequences are known \cite{Newey,Adleman,Galbiati}. 

Moreover, we show that the complexity of the problem in the quantum switch model is $n\log_2(n) +\Theta(n)$. This problem is actually strongly related to the classical problem of permutation networks \cite{Waksman}, i.e. the problem of implementing a permutation in the classical binary circuit model.

\section{Bounds for the standard model}
In \cite{Colnaghi} a simple circuit is provided which solves \upn within $n^2$ calls to the unitary gates. The circuit is made of $n+1$ layers, each composed of the controlled-swaps, interspersed by $n$ $U_i$'s in parallel. 

\vspace{0.1cm}
\centerline{$
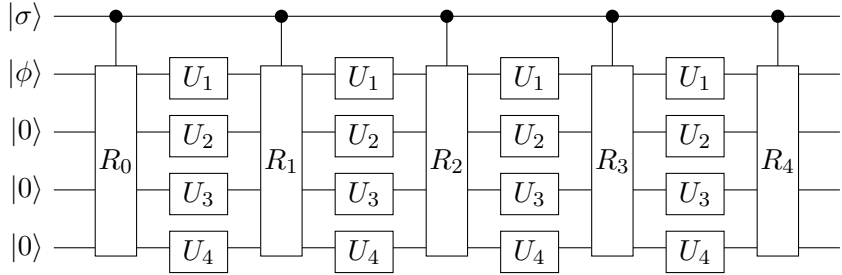
\begin{tikzpicture}[scale=1.1]
\tikzstyle{phase} = [fill,shape=circle,minimum size=5pt,inner sep=0pt];
\foreach \i in {1,...,3} {
  \path[draw] (0,0.7*\i-0.7) node[left]{$\ket 0$}-- (9.5,0.7*\i-0.7);
}
\path[draw] (0,2.1) node[left]{$\ket\phi$} -- (9.5, 2.1);
\path[draw] (0, 2.8) node[left]{$\ket\sigma$} -- (9.5,2.8);
\foreach \i in {0,...,4} {
  \path[draw,fill=white] (0.5+2*\i, 2.2) rectangle node{$R_{\i}$} (1+2*\i, -0.1);
  \path[draw] (0.75+2*\i, 2.2) -- (0.75+2*\i, 2.8) node[phase]{};
}
\foreach \i in {1,...,4} {
 \foreach \j in {1,...,4} {
  \path[draw,fill=white] (1.4+2*\i-2, 2.3-0.7*\j+0.7) rectangle node{$U_{\j}$} (2.1+2*\i-2, 1.8-0.7*\j+0.7);
}
}
\end{tikzpicture}$}
\vspace{0.1cm}

Each layer of generalised controlled-swap ($\Lambda R_i$) performs a rewiring of the qubits: the first layer maps qubit $1$ (the data qubit) to  qubit $\sigma(1)$, on which $U_{\sigma(1)}$ is applied, then the second layer of controlled-swap maps qubit $\sigma(1)$ to qubit $\sigma(2)$ and so on.

In this section we provide a more efficient circuit to solve \upn using the \emph{permutation as substrings} problem introduced by Karp (see \cite{chvataletal}, Problem 36). A sequence over the finite set $[n]$ is complete if it contains all permutations of $[n]$ as a (not necessarily consecutive) subsequence. For instance $w=1213121$ is a complete sequence for $n=3$.  Finding the shortest complete sequence is known as the \emph{permutation as substring} problem. 

\begin{definition}
Given $n\ge 1$, let $S(n)$ be the size of the shortest sequence over $[n]$ which contains each permutation of $[n]$ as a subsequence. 
\end{definition}

Sequences of size $n^2-2n+4$ are known to be complete \cite{Newey,Adleman,Galbiati,Mohanty,Savage}, and $n^2-2n+4$ is actually the size of the shortest complete subsequences for $3\le n \le 7$.  When $n\ge 10$, the size of the shortest complete sequence is upper bounded by $n^2-2n+3$ \cite{Zalinescu}, whereas the best known upper bound for $n\ge 13$ is $\lceil n^2-\frac 7 3 n+\frac{19}3\rceil$ which has been recently established \cite{Radomirovic}. Regarding the lower bound, Kleitman and Kwiatkowski \cite{Kleitman} proved that $S(n)\ge n^2-C_\epsilon n^{7/4+\epsilon}$ for any $\epsilon>0$ where $C_\epsilon$ is a constant depending on $\epsilon$. 

\begin{theorem}
There exists a quantum circuit which solves the \upn problem with $S(n)$ calls. 
\end{theorem}

\noindent{\bf Proof:}
Given a complete sequence $w$ of $[n]$ of size $S(n)$, for any permutation $\sigma$ of $[n]$ let $f_{\sigma}$ be the indices of a subsequence of $w$ which corresponds to $\sigma$, i.e.  $f_\sigma : [n] \to [S(n)]$ is an increasing function s.t. $\forall i\in [n], \sigma(i) = w_{f_\sigma(i)}$. We consider the circuit acting on 3 sub-registers: the control register which contains the description of the permutation $\sigma$, the second register is the data register, and the third one is an auxiliary register initialised in an arbitrary state, say $\ket0$. The circuit is composed of $n+1$ layers of ``controlled-swap'' gates $\Lambda R_0, \ldots , \Lambda R_n$  defined as $\Lambda R_i\ket{\sigma,x,y} = \begin{cases}\ket{\sigma,x,y} &\text{ if $i \in Im(f_\sigma)$}\\\ket{\sigma,y,x} &\text{ otherwise}\end{cases}$ where $Im(f_\sigma) = \{f_\sigma(i) ~|~ i \in [n]\}$ is the image of $f_\sigma$. The unitary transformations $\Lambda R_{i{-}1}$ and $\Lambda R_i$ are interspersed by a call to $U_{w_i}$ on the data register and the identity on the auxiliary register. Given a permutation $\sigma$, the $R_i$ act either as a swap or as the identity in such a way that the unitary transformations applied on the data register are $U_{\sigma(1)}$ then $U_{\sigma(2)}$ and so on.  An example of complete sequence for $n=3$ is $w = 1213121$ which leads to the following circuit:

\vspace{0.3cm}
\centerline{$
\begin{tikzpicture}[scale=1]
\tikzstyle{phase} = [fill,shape=circle,minimum size=5pt,inner sep=0pt];
\foreach \i in {0} {
  \path[draw] (0,0.7*\i) node[left]{$\ket 0$}-- (11.7,0.7*\i);
}
\path[draw] (0,0.7) node[left]{$\ket\phi$} -- (11.7, 0.7);
\path[draw] (0, 1.4) node[left]{$\ket\sigma$} -- (11.7,1.4);
\foreach \i in {0,...,7} {
  \path[draw,fill=white] (0.5+1.5*\i, 0.9) rectangle node{$R_{\i}$} (1+1.5*\i, -0.1);
  \path[draw] (0.75+1.5*\i, 0.9) -- (0.75+1.5*\i, 1.4) node[phase]{};
}
  \path[draw,fill=white] (1.4+1.5*1-1.7, 0.4) rectangle node{$U_{1}$} (2+1.5*1-1.7, 1);
    \path[draw,fill=white] (1.4+1.5*2-1.7, 0.4) rectangle node{$U_{2}$} (2+1.5*2-1.7, 1);
  \path[draw,fill=white] (1.4+1.5*3-1.7, 0.4) rectangle node{$U_{1}$} (2+1.5*3-1.7, 1);
  \path[draw,fill=white] (1.4+1.5*4-1.7, 0.4) rectangle node{$U_{3}$} (2+1.5*4-1.7, 1);
  \path[draw,fill=white] (1.4+1.5*5-1.7, 0.4) rectangle node{$U_{1}$} (2+1.5*5-1.7, 1);
  \path[draw,fill=white] (1.4+1.5*6-1.7, 0.4) rectangle node{$U_{2}$} (2+1.5*6-1.7, 1);
  \path[draw,fill=white] (1.4+1.5*7-1.7, 0.4) rectangle node{$U_{1}$} (2+1.5*7-1.7, 1);
\end{tikzpicture}$}

\vspace{0.3cm}

\begin{corollary}
The \upn problem can be solved within $n^2-2n+4$ calls in the standard model.
\end{corollary}

We conjecture that $S(n)$ is also a lower bound on the number of calls necessary to solve the unitary permutation problem in the standard model, which would imply that any circuit which solves the unitary permutation problem uses at least $n^2- o(n^{7/4+\epsilon})$ queries for any $\epsilon>0$. 

\begin{conjecture}\label{conj}
The \upn problem requires $n^2- o(n^{7/4+\epsilon})$ calls in the standard quantum circuit model, for any $\epsilon>0$. 
\end{conjecture}
We prove the conjecture in a particular setting where only rewiring gates -- like controlled swaps -- are allowed. 

\begin{definition}[Rewiring gates] 
A rewiring gate $R$ is a unitary gate acting on a control register and a $k$-qubit target register as follows: for any  permutation $\sigma$ of $[n]$,  $R\ket{\sigma,x_1, \ldots x_k} = \ket{\sigma, x_{\tau_1}, \ldots x_{\tau_k}}$ where $\tau$ is a permutation of $[k]$ which depends on $\sigma$. 
\end{definition}

\begin{lemma}\label{lowerbound}
Any circuit composed of rewiring gates and calls to the $U_i$ which solves the \upn  problem is composed of at least $n^2-o(n^{7/4+\epsilon})$ calls to the oracle for any $\epsilon>0$.
\end{lemma}

\noindent{\bf Proof.} 
The circuit that solves the \upn problem has 3 inputs: the permutation $\ket \sigma$, the input state $\ket \phi$ and some ancillary qubits that we assume w.l.o.g. in the state $\ket 0$. The  calls to the oracle are performed in a certain order, independent  of the permutation $\sigma$, which can be represented by a sequence $w$ over the set $\{1,\ldots,n\}$: the first call is to $U_{w_1}$, the second to $U_{w_2}$ and so on. If two or more calls are made in parallel we arbitrarily sequentialise the calls. Each call to the oracle is preceded and followed by a rewiring gate (or by the identity which is a particular rewiring gate). Thus for any fixed input permutation $\sigma$, the input state goes through some of the $U$-gates. So the applied unitary is $U_{\tau(m)}\ldots U_{\tau(1)}$ for some sub sequence $\tau$ of $w$ of size $m$, such that $U_{\sigma(n)}\ldots U_{\sigma(1)} = \alpha U_{\tau(m)}\ldots U_{\tau(1)}$ for some $\alpha \in \mathbb C$ (unitary transformations are usually defined up to a global phase).  
We consider the following particular family of unitary gates $U_j$ acting on $2$ registers as follows $\forall d\in \mathbb N,\forall x\in \{0,1\}$,  $U_j \ket{d,x} = e^{ixjn^d}\ket {d+1,x}$.  $U_{\sigma(n)}\ldots U_{\sigma(1)} = \alpha U_{\tau(m)}\ldots U_{\tau(1)}$ implies $U_{\sigma(n)}\ldots U_{\sigma(1)} \ket{0,0}= \alpha U_{\tau(m)}\ldots U_{\tau(1)}\ket{0,0} $  so $\ket{n,0}=\alpha\ket {m,0}$, as a consequence  $\alpha =1$ and $n=m$. Moreover $U_{\sigma(n)}\ldots U_{\sigma(1)} \ket{0,1}=  U_{\tau(n)}\ldots U_{\tau(1)}\ket{0,1}$ $ \implies  e^{i\sum_{d=0}^{n-1}\sigma(d)n^d}\ket{n,1}= e^{i\sum_{d=0}^{n-1}\tau(d)n^d}\ket{n,1}$, thus $\tau = \sigma$.  As a consequence any permutation is a subsequence of $w$, so $w$ is complete and its size, i.e. the number of calls, is at least $n^2 - C_\epsilon n^{7/4+\epsilon}$, for any $\epsilon >0$ \cite{Kleitman}.  So, for any $\epsilon >0$, the minimal number of calls is at least $n^2 - C_{\frac \epsilon 2} n^{7/4+\frac \epsilon 2} = n^2 - o(n^{7/4+\epsilon})$. $\hfill \Box$~\\

Although the unitary permutation problem seems to be strongly related to the permutation as substring problem, conjecture \ref{conj} is false if one considers a slightly different model. For instance, the \upn problem can be solved with a non zero probability using $n$ calls only. Such circuit is based on the teleportation and generalises the construction given for the particular case $n=2$ in \cite{Chiribella}. In a postselected quantum circuit model where one can choose the outcome of each measurement among those which occur with a non zero probability,  the \upn problem can be solved within $n$ calls.  
The probabilistic and postselection settings point out that the proof of conjecture \ref{conj} should rely on some fundamental properties of the quantum circuits, like  causality. In \cite{Chiribella}, the case $n=2$ of the conjecture is proved. The proof is based on the fact that time loops are forbidden in quantum circuits.

\section{Bounds for the quantum switch circuit model}

A quantum switch (\QS)  is a gate that inputs two unitary transforms and a control bit and performs a switch or not of the unitary transformations depending on the control qubit: $\QS (x,U_0,U_1) = (U_x, U_{\bar x})$. Following \cite{Colnaghi}, $\QS$ gate are represented as follows where the control bit is omitted. 
$$
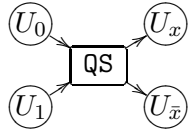
\vcenter{\xymatrix @C=.4em @R=-0.5em @! {
*+[o][F-]{U_0} \ar[dr] & & *+[o][F-]{U_x}  \\
 & *+[F-:<1pt>] {\, \QS\, } \ar[ur] \ar[dr] & \\
*+[o][F-]{U_1} \ar[ur] & & *+[o][F-]{U_{\bar x}}
}}
$$

 In \cite{Colnaghi}, it is proved that the following network (for $n=4$) solves the unitary permutation problem:

$$\vcenter
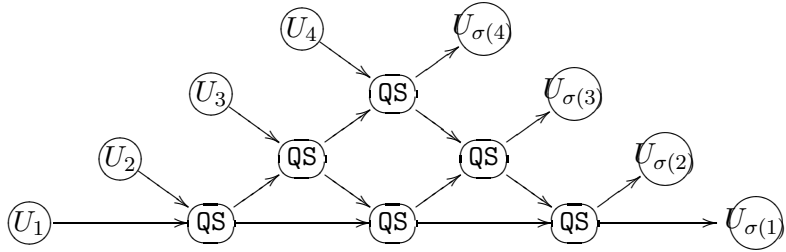
{
\xymatrix @C=.4em @R=-0.5em @! {
& & & *+[o][F-]{U_4} \ar[dr] & & *+[o][F-]{U_{\sigma(4)}} &  \\
& & *+[o][F-]{U_3} \ar[dr] & & *+[F-:<6pt>]{\QS } \ar[dr] \ar[ur] & & *+[o][F-]{U_{\sigma(3)}} &  \\
& *+[o][F-]{U_2} \ar[dr] & & *+[F-:<6pt>]{ \QS } \ar[dr] \ar[ur] & & *+[F-:<6pt>]{ \QS } \ar[dr] \ar[ur] & & *+[o][F-]{U_{\sigma(2)}}   \\
*+[o][F-]{U_1} \ar[rr] & & *+[F-:<6pt>]{ \QS} \ar[ur] \ar[rr] & & *+[F-:<6pt>]{ \QS} \ar[rr] \ar[ur] & & *+[F-:<6pt>]{ \QS } \ar[rr] \ar[ur] & & *+[o][F-]{U_{\sigma(1)}}}}
$$

More generally,  the \upn problems can be solved using $\frac {n(n-1)}2$ quantum switches. Even if this network is claimed to solve the \upn problem in the most efficient way, minimising the number of \QS ~in \cite{Colnaghi}, we show that the problem can be solved much more efficienty  using $n\log_2(n)+O(n)$ \QS ~only. 
To this end, we use  the Bene\v{s} network that solves the classical permutation network problem \cite{Waksman, Benes, Nassimi}.

\begin{lemma}
For any $n\ge 1$ there exists a  circuit composed of $n \log_2(n)  + O(n)$ \emph{\QS} ~which solves the \upn  problem. 
\end{lemma}

\noindent {\bf Proof:} We use the following Bene\v{s} network $B(n)$ to realise any arbitrary permutation \cite{Waksman}: 

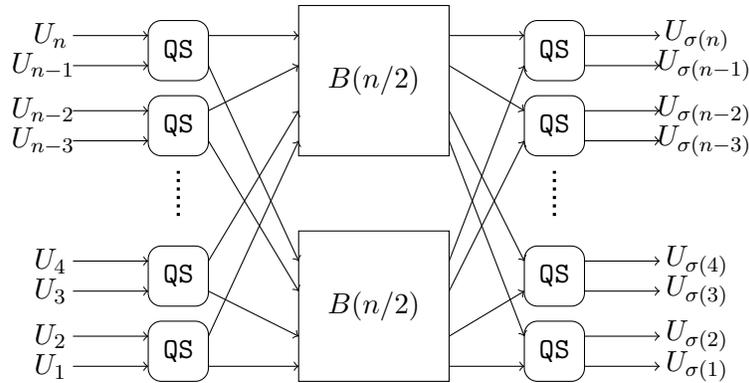
$$
\begin{tikzpicture}
\draw (0,0) rectangle node{$B(n/2)$} (2,2);
\draw (0,3) rectangle node{$B(n/2)$} (2,5);
\draw[rounded corners=5pt] (-2,0) rectangle node{\QS} (-2+0.8,0.8);
\draw[rounded corners=5pt] (-2,1) rectangle node{\QS} (-2+0.8,1+0.8);
\path[draw,dotted,very thick] (-1.6,2.2) -- (-1.6,2.8);
\draw[rounded corners=5pt] (-2,3) rectangle node{\QS} (-2+0.8,3+0.8);
\draw[rounded corners=5pt] (-2,4) rectangle node{\QS} (-2+0.8,4+0.8);
\draw[rounded corners=5pt] (3,0) rectangle node{\QS} (3+0.8,0.8);
\draw[rounded corners=5pt] (3,1) rectangle node{\QS} (3+0.8,1+0.8);
\path[draw,dotted,very thick] (3.4,2.2) -- (3.4,2.8);
\draw[rounded corners=5pt] (3,3) rectangle node{\QS} (3+0.8,3+0.8);
\draw[rounded corners=5pt] (3,4) rectangle node{\QS} (3+0.8,4+0.8);
\path[draw,->] (-3,0.2) -- (-2,0.2);
\path[draw,->] (-3,0.6) -- (-2,0.6);
\path[draw,->] (-3,1+0.2) -- (-2,1+0.2);
\path[draw,->] (-3,1+0.6) -- (-2,1+0.6);
\path[draw,->] (-3,3+0.2) -- (-2,3+0.2);
\path[draw,->] (-3,3+0.6) -- (-2,3+0.6);
\path[draw,->] (-3,4+0.2) -- (-2,4+0.2);
\path[draw,->] (-3,4+0.6) -- (-2,4+0.6);
\path[draw,->] (3.8,0.2) -- (4.8,0.2);
\path[draw,->] (3.8,0.6) -- (4.8,0.6);
\path[draw,->] (3.8,1+0.2) -- (4.8,1+0.2);
\path[draw,->] (3.8,1+0.6) -- (4.8,1+0.6);
\path[draw,->] (3.8,3+0.2) -- (4.8,3+0.2);
\path[draw,->] (3.8,3+0.6) -- (4.8,3+0.6);
\path[draw,->] (3.8,4+0.2) -- (4.8,4+0.2);
\path[draw,->] (3.8,4+0.6) -- (4.8,4+0.6);
\path[draw,->] (-1.2,0.2) -- (0,0.2);
\path[draw,->] (-1.2,0.6) -- (0,3+0.2);
\path[draw,->] (-1.2,1+0.2) -- (0,0.6);
\path[draw,->] (-1.2,1+0.6) -- (0,3+0.6);
\path[draw,->] (-1.2,3+0.2) -- (0,1+0.2);
\path[draw,->] (-1.2,3+0.6) -- (0,3+1+0.2);
\path[draw,->] (-1.2,3+1+0.2) -- (0,1+0.6);
\path[draw,->] (-1.2,3+1+0.6) -- (0,3+1+0.6);
\path[draw,->] (2,0.2) -- (3,0.2);
\path[draw,->] (2,0.6) -- (3,1+0.2);
\path[draw,->] (2,1+0.2) -- (3,3+0.2);
\path[draw,->] (2,1+0.6) -- (3,3+1+0.2);
\path[draw,->] (2,3+0.2) -- (3,0.6);
\path[draw,->] (2,3+0.6) -- (3,1+0.6);
\path[draw,->] (2,3+1+0.2) -- (3,3+0.6);
\path[draw,->] (2,3+1+0.6) -- (3,3+1+0.6);
\node at (-3.3,0.2) {$U_1$};
\node at (-3.3,0.6) {$U_2$};
\node at (-3.3,1.2) {$U_3$};
\node at (-3.3,1.6) {$U_4$};
\node at (-3.4,3.2) {$U_{n-3}$};
\node at (-3.4,3.6) {$U_{n-2}$};
\node at (-3.4,3+1.2) {$U_{n-1}$};
\node at (-3.3,3+1.6) {$U_n$};
\node at (5.3,0.2) {$U_{\sigma(1)}$};
\node at (5.3,0.6) {$U_{\sigma(2)}$};
\node at (5.3,1.2) {$U_{\sigma(3)}$};
\node at (5.3,1.6) {$U_{\sigma(4)}$};
\node at (5.4,3+0.2) {$U_{\sigma(n-3)}$};
\node at (5.4,3+0.6) {$U_{\sigma(n-2)}$};
\node at (5.4,3+1.2) {$U_{\sigma(n-1)}$};
\node at (5.3,3+1.6) {$U_{\sigma(n)}$};
\end{tikzpicture}
$$

\noindent The size of the Bene\v{s} network is smaller than $n(\log_2(n) -\frac12)$ even if $n$ is not a power of $2$ (see \cite{chang} for details on Bene\v{s} networks when $n$ is not a power of $2$).  $\hfill \Box$~\\

The previous circuit is optimal, indeed any circuit which solves the \upn~problem is composed of at least $n\log_2(n) -2n$ \QS~gates: 

\begin{lemma}
For any $n\ge 1$, a QS circuit solving the \upn~problem is composed of at least $\lceil \log_2(n!) \rceil  \ge n\log_2(n)-2n$ quantum switches.
\end{lemma}

\noindent {\bf Proof.} A circuit composed of $k$ quantum switches can produce at most  $2^k$ different orderings of the $U_i$. As the $U_i$ can be chosen such that for every distinct permutations $\sigma, \tau$, $U_{\sigma(n)}\ldots U_{\sigma(1)} \neq  U_{\tau(n)}\ldots U_{\tau(1)}$ (see proof of Lemma \ref{lowerbound}), $2^k$ must be larger than $n!$ the number of possible permutations. So $k\ge \lceil \log_2(n!)\rceil \ge n\log_2(n)-\frac{n-1}{\ln(2)}\ge n\log_2(n)-2n$.\hfill$\Box$

\end{document}

%% file: myQcircuit.tex
\usepackage[matrix,frame,arrow]{xy}
\usepackage{amsmath}

\newcommand{\ket}[1]{\left\vert{#1}\right\rangle}
\newcommand{\node}[2][]{{\begin{array}{c} \ _{#1}\  \\ {#2} \\ \ \end{array}}\drop\frm{o} }





%% file: UnitaryPermutations-arxiv.bbl
\begin{thebibliography}{10}
\providecommand{\bibitemdeclare}[2]{}
\providecommand{\surnamestart}{}
\providecommand{\surnameend}{}
\providecommand{\urlprefix}{Available at }
\providecommand{\url}[1]{\texttt{#1}}
\providecommand{\href}[2]{\texttt{#2}}
\providecommand{\urlalt}[2]{\href{#1}{#2}}
\providecommand{\doi}[1]{doi:\urlalt{http://dx.doi.org/#1}{#1}}
\providecommand{\bibinfo}[2]{#2}

\bibitemdeclare{article}{Adleman}
\bibitem{Adleman}
\bibinfo{author}{L.~\surnamestart Adleman\surnameend} (\bibinfo{year}{1974}):
  \emph{\bibinfo{title}{Short permutation strings}}.
\newblock {\sl \bibinfo{journal}{Discrete Mathmatics}} \bibinfo{volume}{10}, p.
  \bibinfo{pages}{197}.

\bibitemdeclare{book}{Benes}
\bibitem{Benes}
\bibinfo{author}{V.~\surnamestart Bene\v{s}\surnameend} (\bibinfo{year}{1965}):
  \emph{\bibinfo{title}{Mathematical Theory of Connecting Networks and
  Telephone Traffic}}.
\newblock \bibinfo{publisher}{Academic Press}.

\bibitemdeclare{article}{chang}
\bibitem{chang}
\bibinfo{author}{C.~\surnamestart Chang\surnameend} \&
  \bibinfo{author}{R.~\surnamestart Melhem\surnameend} (\bibinfo{year}{1997}):
  \emph{\bibinfo{title}{Arbitrary Size Benes Networks}}.
\newblock {\sl \bibinfo{journal}{Parallel Processing Letters}}
  \bibinfo{volume}{7}, pp. \bibinfo{pages}{279--284}.

\bibitemdeclare{article}{Chiribella}
\bibitem{Chiribella}
\bibinfo{author}{G.~\surnamestart Chiribella\surnameend},
  \bibinfo{author}{G.~M. \surnamestart D'Ariano\surnameend},
  \bibinfo{author}{P.~\surnamestart Perinotti\surnameend} \&
  \bibinfo{author}{B.~\surnamestart Valiron\surnameend} (\bibinfo{year}{2013}):
  \emph{\bibinfo{title}{Quantum computations without definite causal
  structure}}.
\newblock {\sl \bibinfo{journal}{Physical Review A}} \bibinfo{volume}{88}, p.
  \bibinfo{pages}{022318}.

\bibitemdeclare{article}{chvataletal}
\bibitem{chvataletal}
\bibinfo{author}{V.~\surnamestart Chv\'atal\surnameend}, \bibinfo{author}{D.~A.
  \surnamestart Klarner\surnameend} \& \bibinfo{author}{D.~E. \surnamestart
  Knuth\surnameend} (\bibinfo{year}{1972}): \emph{\bibinfo{title}{Selected
  combinatorial research problems. Technical Report 292}}.

\bibitemdeclare{article}{Colnaghi}
\bibitem{Colnaghi}
\bibinfo{author}{T.~\surnamestart Colnaghi\surnameend}, \bibinfo{author}{G.~M.
  \surnamestart D'Ariano\surnameend}, \bibinfo{author}{S.~\surnamestart
  Facchini\surnameend} \& \bibinfo{author}{P.~\surnamestart
  Perinotti\surnameend} (\bibinfo{year}{2012}): \emph{\bibinfo{title}{Quantum
  computation with programmable connections between gates}}.
\newblock {\sl \bibinfo{journal}{Physics Letters A}} \bibinfo{volume}{376}, p.
  \bibinfo{pages}{2940}.

\bibitemdeclare{article}{Galbiati}
\bibitem{Galbiati}
\bibinfo{author}{G~\surnamestart Galbiati\surnameend} \& \bibinfo{author}{F.~P.
  \surnamestart Preparata\surnameend} (\bibinfo{year}{1976}):
  \emph{\bibinfo{title}{On permutation embedding sequences}}.
\newblock {\sl \bibinfo{journal}{SIAM Journal of Applied Mathematics}}
  \bibinfo{volume}{30}, p. \bibinfo{pages}{421}.

\bibitemdeclare{article}{Kleitman}
\bibitem{Kleitman}
\bibinfo{author}{D.~\surnamestart Kleitman\surnameend} \&
  \bibinfo{author}{D.~\surnamestart Kwiatkowsky\surnameend}
  (\bibinfo{year}{1976}): \emph{\bibinfo{title}{A lower bound on the length of
  a sequence containing all permutations as subsequences}}.
\newblock {\sl \bibinfo{journal}{Journal of Combinatorial Theory Series A}}
  \bibinfo{volume}{21}, p. \bibinfo{pages}{129}.

\bibitemdeclare{article}{Mohanty}
\bibitem{Mohanty}
\bibinfo{author}{S.~P. \surnamestart Mohanty\surnameend}
  (\bibinfo{year}{1980}): \emph{\bibinfo{title}{Shortest string containing all
  permutations}}.
\newblock {\sl \bibinfo{journal}{Discrete Mathematics}} \bibinfo{volume}{31},
  p.~\bibinfo{pages}{91}.

\bibitemdeclare{article}{Nassimi}
\bibitem{Nassimi}
\bibinfo{author}{D.~\surnamestart Nassimi\surnameend} \&
  \bibinfo{author}{S.~\surnamestart Sahni\surnameend} (\bibinfo{year}{1981}):
  \emph{\bibinfo{title}{A Self-Routing Benes Network and Parallel Permutation
  Algorithms}}.
\newblock {\sl \bibinfo{journal}{IEEE Transactions on Computers}}
  \bibinfo{volume}{C-30}, p. \bibinfo{pages}{332}.

\bibitemdeclare{article}{Newey}
\bibitem{Newey}
\bibinfo{author}{M.~\surnamestart Newey\surnameend} (\bibinfo{year}{1973}):
  \emph{\bibinfo{title}{Notes on a problem involving permutations as
  subsequences. Technical Report 340}}.

\bibitemdeclare{article}{Radomirovic}
\bibitem{Radomirovic}
\bibinfo{author}{S.~\surnamestart Radomirovi\'c\surnameend}
  (\bibinfo{year}{2012}): \emph{\bibinfo{title}{A Construction of Short
  Sequences Containing All Permutations of a Set as Subsequences}}.
\newblock {\sl \bibinfo{journal}{The electronic journal of combinatorics}}
  \bibinfo{volume}{19}, p.~\bibinfo{pages}{31}.

\bibitemdeclare{article}{Savage}
\bibitem{Savage}
\bibinfo{author}{C.~\surnamestart Savage\surnameend} (\bibinfo{year}{1982}):
  \emph{\bibinfo{title}{Short strings containing all k-element permutations}}.
\newblock {\sl \bibinfo{journal}{Discrete Mathematics}} \bibinfo{volume}{42},
  p. \bibinfo{pages}{281}.

\bibitemdeclare{article}{Waksman}
\bibitem{Waksman}
\bibinfo{author}{A.~\surnamestart Waksman\surnameend} (\bibinfo{year}{1968}):
  \emph{\bibinfo{title}{A permutation network}}.
\newblock {\sl \bibinfo{journal}{J. Ass. Comput. Mach.}} \bibinfo{volume}{15},
  p. \bibinfo{pages}{159}.

\bibitemdeclare{article}{Zalinescu}
\bibitem{Zalinescu}
\bibinfo{author}{E.~\surnamestart Z\u{a}linescu\surnameend}
  (\bibinfo{year}{2011}): \emph{\bibinfo{title}{Shorter strings containing all
  k-element permutations}}.
\newblock {\sl \bibinfo{journal}{Information Processing Letters}}
  \bibinfo{volume}{111}, p. \bibinfo{pages}{605}.

\end{thebibliography}
